\documentclass[11pt]{article}

\usepackage{geometry}                
\usepackage{graphicx}
\newcommand{\be}{\begin{equation}}
\newcommand{\ee}{\end{equation}}
\newcommand{\bea}{\begin{eqnarray}}
\newcommand{\eea}{\end{eqnarray}}
\newcommand{\beas}{\begin{eqnarray*}}
\newcommand{\eeas}{\end{eqnarray*}}
\newcommand{\avg}[1]{\left\langle{#1}\right\rangle}

\usepackage{amssymb}
\usepackage{amsmath}
\usepackage{amsfonts}
\usepackage{graphicx}
\usepackage{epstopdf}

\begin{document}

\title{On information efficiency and financial stability}
\author{Fabio Caccioli$^{1}$ \& Matteo Marsili$^2$
~~~~~\\
{\em 1 - SISSA, Via Beirut 2-4, 34151 Trieste, Italy}\\ 
{\em 2 - Abdus Salam International Centre for Theoretical Physics,}\\
{\em  Strada Costiera 11, 34151 Trieste, Italy}\\ 
}
\maketitle
\begin{abstract}
We study a simple model of an asset market with informed and non-informed agents. In the absence of non-informed agents, the market becomes information efficient when the number of traders with different private information is large enough. Upon introducing non-informed agents, we find that the latter contribute significantly to the trading activity if and only if the market is (nearly) information efficient. This suggests that information efficiency might be a necessary condition for bubble phenomena -- induced by the behavior of non-informed traders -- or conversely that throwing some sands in the gears of financial markets may curb the occurrence of bubbles. 
\end{abstract}

\section{Introduction}

Financial markets have increased tremendously in size and complexity in the last decades, with the proliferation of hedge funds and the expansion of derivative markets. Within the neo-classical paradigm, the expansion in the diversity of traders and  in the repertoire of financial instruments is, generally, enhancing the efficiency of the market (see however \cite{Hart75}, \cite{CassCitanna}). Indeed unfettered access to trading in financial markets makes more liquidity available and it eliminates arbitrages, thus pushing the market closer to the theoretical limit of {\em perfectly competitive, informationally efficient markets}. Likewise, the expansion in the repertoire of trading instruments provides a wider range of possibilities to hedge risks and it drives the system closer to the theoretical limit of {\em dynamically complete markets} \cite{MertonBodie}. Both conclusions rely on non-trivial assumptions, notably the absence of information asymmetries. Indeed, financial stability is related to the effects of asymmetric information and most of the responsibility for market failures is, in one way or another, usually put on market imperfections\footnote{According to F.S. Mishkin "Financial instability occurs when there is a disruption to financial markets in which asymmetric 
information and hence adverse selection and moral hazard problems become much worse, so that
financial markets are unable to channel funds efficiently to those with the most productive
investment opportunities." \cite{Iceland,Mishkin}. For an account of perverse effect which led to the 2007 -- 2008 crisis in credit markets, see for example \cite{Jarrow}.}.
Market imperfections are inevitable even in stable periods, so a relevant question is to  understanding when deviations from ideal conditions are amplified by the internal dynamics of the market, leading to a full blown crisis. 

This paper suggests that the more markets are close to ideal conditions, the more they are prone to the proliferation and amplification of market imperfections. This point has already been made in the literature \cite{BHW2006,CMV2009,MM2009} concerning the expansion in the repertoire of financial instruments\footnote{Specifically, \cite{BHW2006} show that adding more and more Arrow's securities in a market with heterogeneous adaptive traders, brings the system to a dynamic instability. A similar conclusion was drawn in  \cite{CMV2009}, though based on different models.  \cite{MM2009} discusses instead an equilibrium model and it shows that as the number of possible trading instruments increases the market approaches the theoretical limit of complete markets but allocations develop a marked sensitivity to price indeterminacy, and the volume of trading implied by hedging in the interbank market diverges.}.

Here we address the issue of stability, in relation to information efficiency. Information efficiency refers to the ability of the market to allocate investment to activities which provide profitable return opportunities. In brief, traders who have a private information on the performance of an asset will buy or sell shares of the corresponding stock in order to make a profit. As a result, prices will move in order to incorporate this information, thus reducing the profitability of that piece of information. In equilibrium, when all informed traders are allowed to invest, prices must be such that no profit can be extracted from the market. 

In this respect, markets behave as information processing and aggregating devices and, in the ideal limit, market prices are expected to reflect all possible information: this is the content of the celebrated {\em Efficient Market Hypothesis} \cite{fama}. 
Paradoxically, however, when markets are really informationally efficient, traders have no incentive to gather private information, because prices already convey all possible information.
Hence, as realized long ago \cite{GrossmanStiglitz}, traders' behavior does not transfer any information into prices, which implies that efficient markets cannot be realized. 

The interplay between informed and non-informed traders is one of the key elements in explaining market dynamics. Informed traders , so-called fundamentalists, typically have a stabilizing effect whereas non-informed traders, e.g. trend-followers  or  chartists in general, can destabilize the market and induce bubble phenomena. 
Research in Heterogeneous Agents Models \cite{Hommes,LM99} has provided solid support to the thesis that when trading activity is dominated by non-informed traders, bubbles and instabilities develop. 

Our goal, here, is to establish a relation between market efficiency and
the interplay between informed and non-informed traders. Specifically, we provide support to the idea that {\em non-informed traders dominate if and only if the market is sufficiently close to information efficiency}. In addition, as markets become informationally efficient, they develop a marked susceptibility to perturbations and instabilities. 

Our discussion steps from the simple asset market model studied in \cite{BMRZ}, which describes a population of heterogeneous individuals, who receive a private signal on the return, or dividend,  of a given asset. Given their private signal, agents invest in the asset and their demand determines the price, via market clearing. 
Agents learn how to optimally exploit their private information in their trading activity, which has the effect that their information is incorporated into prices. As the number $N$ of agents with a different private information increases, prices gradually converge to the returns. Beyond a critical number of agents, prices converge exactly to returns. Therefore, the model provides a stylized picture of how markets aggregate information into prices and become informationally efficient. 

In this setting, we introduce non-informed agents who adopt the same learning dynamics, but which base their decision on public information, rather than on a private signal. In particular, we take the sign of the last return as public signal, which mimics a chartist behavior, in its simplest form. Our main result is that chartists take over a sizable share of market activity only when the market becomes informationally efficient.

The rest of the paper is organized as follows. The next section introduces the model and the notation. In the following section we first recall the results with only informed traders and then discuss the effect of introducing non-informed traders. The paper concludes with a discussion of the extension and relevance of the results.

\section{The model: Information efficiency and chartists}

Let us consider a market where a single asset is being traded an infinite number of periods. 
Let there be $N$ informed traders (fundamentalists) and $N'$ uninformed traders (chartists) operating in the market. For simplicity, we assume that all uninformed traders adopt the same trading strategy, so that the description of chartists can be given in terms of a single representative agent. We thus set $N'=1$ and we shall refer to the chartist representative agent as agent $i=0$.

In the market there are $N$ units of asset available at each time and at the end of each period the asset pays a return. 
The return depends only on the state of nature in that period, $\omega=1,\ldots,\Omega$, and is
denoted by $R^{\omega}$. The state of nature is determined, in each
period, independently according to the uniform distribution on
the integers $1,\ldots,\Omega$. 

Traders do not observe the state directly, but informed traders ($i=1,\ldots,N$) receive a signal on the
state according to some fixed private information structure, which is 
determined at the initial time and remains fixed. More precisely, 
a signal is a function from the state $\omega$ to
a signal space, which for simplicity we assume to be $M = \{-,
+\}$. We denote by  $k_i^
\omega\in M$ the signal observed by trader $i$ in state $\omega$. The information structure available to agent $i$ is then encoded in the vector $(k_i^\omega)_{\omega \in \Omega}$. Trader $i=0$, instead, does not receive any signal on the state $\omega$, but she observes a public variable $k_0\in \{-1,+1\}$, such as the sign of the last excess return: ${\rm sign} (R^{\omega}-p^{\omega})$.

We focus on a random realization of this setup, where the
value of the return $R^{\omega}$ in state $\omega$ is drawn at random
before the first period, and does not change afterwards.  Returns thus only
change because the state of nature changes. As in \cite{BMRZ} we take $R^\omega$ Gaussian with mean $\bar R$ and variance $s^2/N$. 
Likewise, the information structure is determined by
setting $k_i^\omega =+1$ or $-1$ with equal probability, 
independently across traders $i$ and states $\omega$. Appendix \ref{infoapp} discusses the information content of signals in more detail.

At the beginning of each period, a state $\omega$ and a public information $k_0$ are drawn, and private information $k_i^\omega$ is revealed to informed agents ($i>0$). All traders decide to invest a monetary amount $z_i^m$ in the asset: here $m\in M$, for $i>0$, takes the value  of the signals $k_i^{\omega}$ which agent $i>0$ receives, whereas it equals $k_0$ for $i=0$. The price of the asset $p^{\omega,k_0}$ is then derived from the market clearing condition
\be
Np^{\omega,k_0}=\sum_{i=1}^N\sum_{m=\pm 1} z_i^{m}\delta_{k_i^{\omega},m}+\sum_{m=\pm1} z_0^{m}\delta_{k_0,m},
\label{price}
\ee
where $\delta_{i,j}=1$ if $i=j$ and $\delta_{i,j}=0$ otherwise.
Agents do not know the price at which they will buy the asset when they decide their investment $z_i^m$. 
The price depends on the
state $\omega$ and on $k_0$ because the amount invested by each agent depends on the signal they receive, which depends on $\omega$ \cite{ShapleyShubik,Pliska}.
At the end of the period, each unit of asset pays a monetary amount
$R^\omega$. If agent $i$ has invested $z_i^m$ units of money, he 
will hold $z_i^m/p^{\omega,k_0}$ units of asset, so the expected payoff of agents is
\be
u_i(z_i)=\frac{1}{\Omega}\sum_\omega\sum_m\delta_{k_i^{\omega},m}z_i^m\left(\frac{R^{\omega}}{p^{\omega,k_0}}-1\right).
\ee

How will agents choose their investments? One can consider either competitive equilibria or take  a dynamical approach where agents are assumed to learn over time how to invest optimally. As in  \cite{BMRZ} these two different choices are going to bring to the same equilibria, so we focus on the latter. 
In particular, each informed agent $i>0$ has a propensity to invest $U_i^m(t)$ for each
of the signals $m=\pm 1$. His investment $z_i^m=\chi(U_i^m)$ at time
$t$ is an increasing function of $U_i^m(t)$ ($\chi: \mathbb{R}\to \mathbb{R}^+$) with
$\chi(x)\to 0$ if $x\to -\infty$ and $\chi(x)\to \infty$ if $x\to
\infty$. After each period agents update $U_i^m(t)$ according to the
marginal success of the investment:
\begin{equation}
U_i^m(t+1)=U_i^m(t)+\left(R^{\omega_t}-p^{\omega_t,k_0}_t\right)\delta_{k_i^{\omega_t},m}
-\frac{\epsilon}{N},\qquad i=1,\ldots,N
\label{learn}
\end{equation}
where $\omega_t$ is the state at time $t$ and $p^{\omega_t,k_0}_t$ is the realized price.
The idea in Eq. (\ref{learn}) is that if for a given signal $m$ agent $i$ observes returns $R^\omega$ which are higher that prices, she will increase her propensity $U_i^m$ to invest under that signal. At odds with  \cite{BMRZ}, the learning dynamics for informed agents ($i>0$) also takes into account the cost of information, through the term $\epsilon$. More precisely, investment is considered attractive ($U_i^m>0$) only if the returns under signal $m$ exceed prices by more than $\epsilon$. 
Similarly, the non-informed agent updates her propensity to trade according to
\begin{equation}
U_0^m(t+1)=U_0^m(t)+\left(R^{\omega_t}-p^{\omega_t,k_0}_t\right)\delta_{k_0,m}
\label{learn0}
\end{equation}
and invests an amount $z_0^m=\chi(U_0^m)$, depending on the value $m=k_0$ of public information at time $t$.  
\section{Results}

\begin{figure}
  \centering
  \includegraphics[width=8cm]{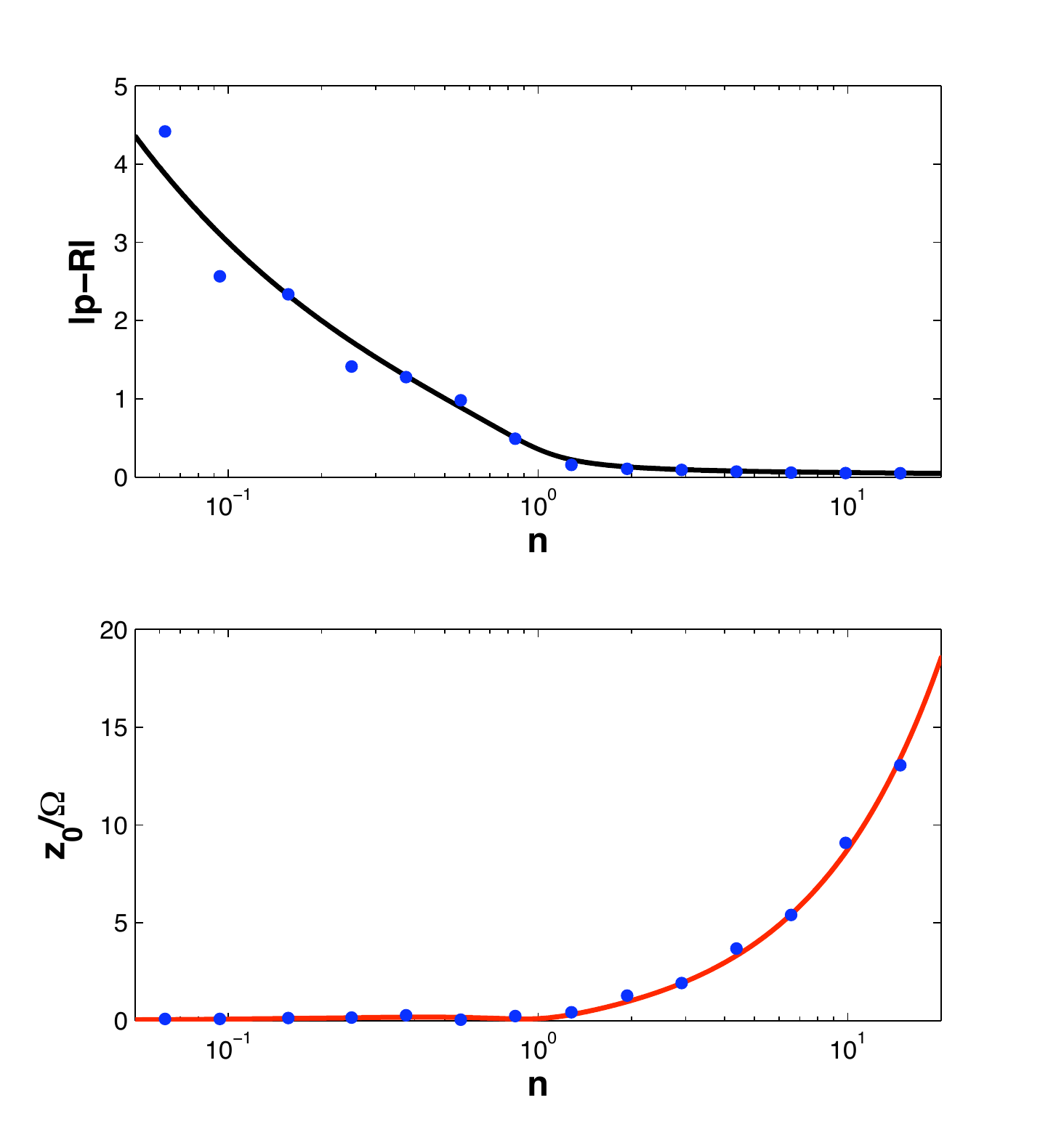}\\
  \caption{\footnotesize \textit{Top panel: distance $|p-R|=\sqrt{\sum_{\omega=1}^\Omega 
E_{k_0}\left[R^\omega-p^{\omega,k_0}\right]^2}$ of prices from returns in competitive equilibrium.  The full line represents the analytical solution for the case $s=\overline{R}=1$ and $\epsilon=0.1$,  points refer to numerical simulations of systems with $\Omega=32$, $s=\overline{R}=1$ and $\epsilon=0.1$.  Bottom panel: monetary amount invested by the trend follower $z_0$ for the same values of the parameters.}}\label{eff1}
\end{figure}
\begin{figure}
\centering
\includegraphics[width=8cm]{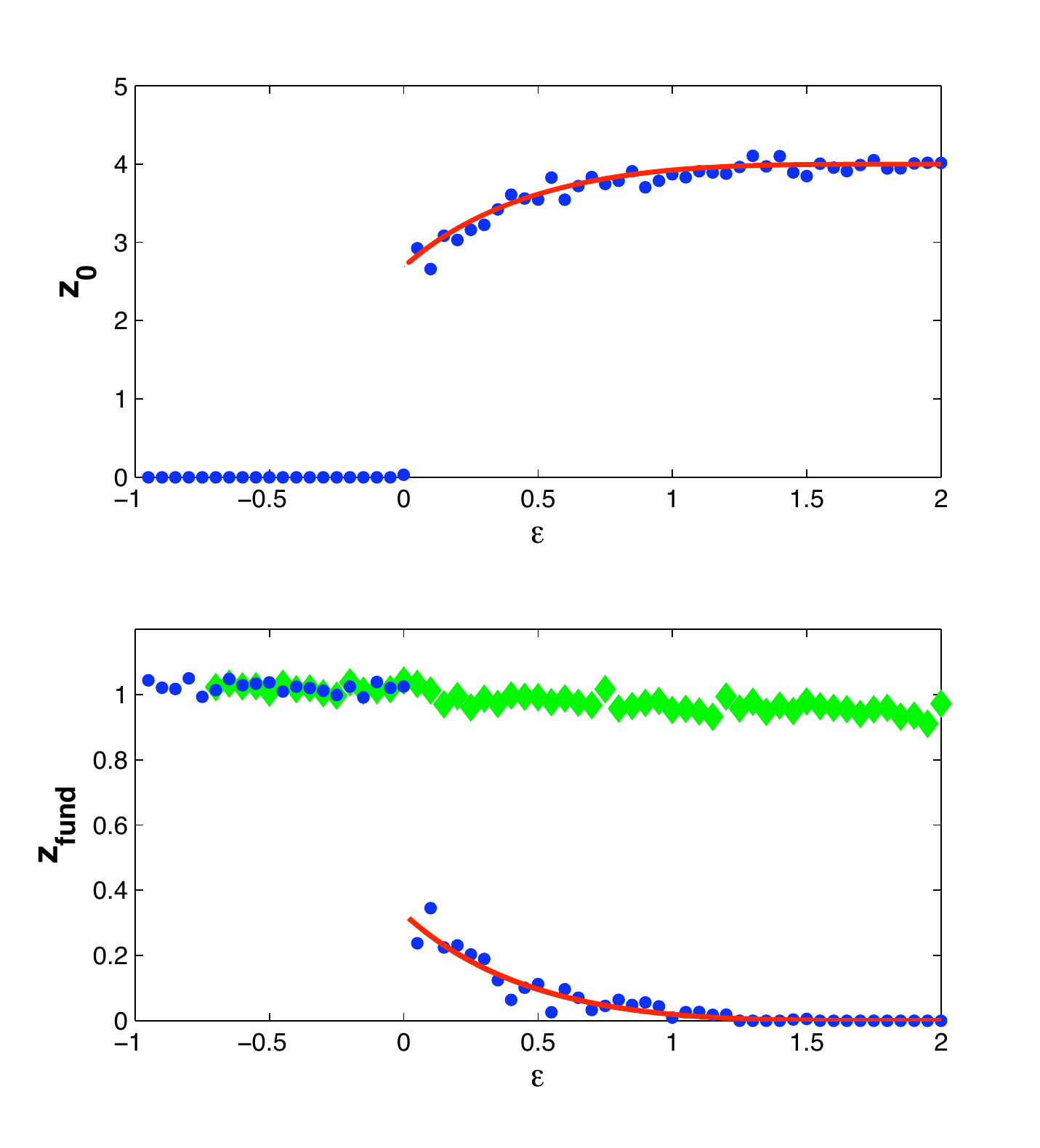}
\caption{\footnotesize {\textit{Top panel: monetary amount invested by the trend follower. Bottom panel:  monetary amount invested by a fundamentalist in presence (blue points) or absence (green diamonds) of the trend follower.  Points refer to simulations of systems with $\Omega=32$, $n=4$ and $s=\overline{R}=1$. Full lines represent the corresponding analytical solution. }}} \label{eff2}
\end{figure}
In this  section we study the typical properties of the market introduced above in the limit $N\to\infty$, $\Omega\to\infty$ and $n=\frac{N}{\Omega}$ finite, where the characterisation of the system can be achieved through a statistical mechanics approach.\\
Let us briefly recall the behavior of the market in the absence ($z_0=0$) of non-informed traders and $\epsilon=0$.  \cite{BMRZ} show that the learning dynamics converges to the allocations $\{z_i^m\}$ which correspond to the solution of the minimization of the function

\begin{equation}
H=\frac{1}{2}\sum_{\omega=1}^\Omega \left(R^\omega-p^\omega\right)^2,\qquad
p^{\omega}=\frac{1}{N}\sum_{i=1}^N\sum_{m=\pm 1} z_i^{m}\delta_{k_i^{\omega},m}.
\label{H}
\end{equation}
The function $H$ is the squared distance of prices from returns. As more and more different types of informed agents enter the market prices approach returns. There is a critical value $n_c$ of informed traders beyond which $H=0$, which implies that prices equal returns ($p^\omega=R^\omega$) for each state $\omega=1,\ldots,\Omega$. As discussed in Appendix \ref{infoapp}, this corresponds to the strong form of market information efficiency, when all private information is incorporated into prices. 

The region $H=0$ is also characterized by a divergent {\em susceptibility}, which means that allocations $\{z_i^m\}$ have a marked dependence on structural parameters. The susceptibility $\Phi$ relates a small uncertainty in a structural parameter, such as e.g. $R^\omega$, to the uncertainty in allocations $\delta z_i^m\simeq \Phi \delta R^\omega$. A divergent susceptibility $\Phi\to\infty$ signals the fact that equilibria with different allocations are possible even for the same structural parameters, i.e. that the minimum of Eq. (\ref{H}) is not unique.

What happens when we introduce chartists ($z_0>0$) and information costs ($\epsilon>0$)? 
First we find that allocations $\{z_i^m\ge 0\}_{i=0,\ldots,N}$ are again given by the solution of the minimization of a function, which takes the form

\begin{equation}
H_\epsilon=\frac{1}{2}\sum_{\omega=1}^\Omega 
E_{k_0}\left[R^\omega-p^{\omega,k_0}\right]^2+\frac{\epsilon}{2 N}\sum_{i=1}^N\sum_{m=\pm 1} z_i^m,
\label{Heps}
\end{equation}
where $p^{\omega,k_0}$ is given in Eq. (\ref{price}) in terms of $z_i^m$, $i=0,\ldots,N$, $m=\pm 1$. The proof proceeds, on one side, by taking the partial derivatives of $H_\epsilon$ with respect to $z_i^m$ and analyzing the Kuhn-Tucker conditions for the minimization of $H_\epsilon$. These tell us that if  the partial derivative of $H_\epsilon$ vanishes in the minimum, then $z_i^m>0$. Otherwise, if the derivative is positive, then $z_i^m=0$. On the other side, one easily finds that 
\begin{equation}
\label{Hepspart}
\frac{\partial H_\epsilon}{\partial z_i^m}=-E_{k_0,\omega}\left[U_i^m(t+1)-U_i^m(t)\right]
\end{equation}
which implies that the Kuhn-Tucker conditions correspond exactly to the conditions for the stationary state (with $U_i^m\to -\infty$ when $z_i^m=0$).

This result paves the way for the extension of the statistical mechanics approach to this case. Some simple heuristic arguments can be useful in order to understand the basic behaviour of the system. Let us consider the case of small $\epsilon$ and small $n$. Then the first term in Eq. (\ref{Heps}) dominates the second and the minimum is expected to be close to that without chartists. When $n$ increases, however, the value of $H$ decreases making the two terms comparable. When this happens, i.e. when $n\approx n_c$ and $H\approx 0$, then it starts to become possible to achieve a small value of $H_\epsilon$ by decreasing the size of the second term increasing, at the same time, $z_0^m$ in order to keep average prices of the same order of average returns. Hence we expect $z_0^m$ to be large and of order $N$ when the market becomes close to information efficient. 
The results of numerical simulations as well as the analytical solution for competitive market equilibrium (see Appendix \ref{calc} for more details), shown in Fig. \ref{eff1}, confirm this picture. Upon increasing the number of informed agents, the system undergoes a transition from inefficient to efficient  market.  Correspondingly, the share of trades due to uninformed agent starts raising only once information has been aggregated by informed traders. It has to be noticed that the introduction of the information cost $\epsilon$ makes sure that a perfect efficiency of the market is recovered only at $\epsilon=0$. It is then instructive to look at the behaviour of the chartists as a function of $\epsilon$. Figure \ref{eff2} shows signatures of a phase transition occurring at $\epsilon=0$.  Indeed, for $\epsilon<0$ the chartists barely operates in the market, while they start trading as soon as $\epsilon>0$. 

\section{Conclusions}

We have shown that, in a simple asset market model, non-informed traders contribute a non-negligible fraction of the trading activity only when the market becomes informationally efficient. In the simple setting studied here, non-informed traders do not have a destabilizing effect on the market as in the models of  \cite{Hommes}. At the same time, when non-informed traders dominate, their activity does not spoil information efficiency.
Nevertheless, we can see from our analysis that information efficiency is associated with a phase transition in the statistical mechanics sense, characterised by strong fluctuations and sharp discontinuities in the optimal allocations. This suggests that market efficiency carries in fact some seeds of instability.\\
Moreover, when combined with the insights of the literature on Heterogeneous Agent Models \cite{Hommes}, the very fact that  non-informed traders start trading massively when market efficiency is approached in fact suggests that information efficiency can trigger the occurrence of bubbles and instabilities. This issue has been also addressed in \cite{goldbaum} recently, however the analysis was limited to a single type of fundamentalist ($N=1$ in our case) and one type of trend followers. 
A stronger case would require first to extend the framework of  \cite{Hommes} to the case of fundamentalists with many different types of private information, recovering a picture for information efficiency similar to that provided by  \cite{BMRZ}. Then one should investigate the effect of introducing non-informed traders, i.e. genuine trend-followers. Besides understanding whether information efficiency is also in that case a necessary condition for non-informed traders to dominate, one could also address the interesting question of the effect of chartists on information efficiency. 

Ultimately, our results suggest that excessive insistence on information efficiency in market regulation policies , as e.g. in the debate on the Tobin tax \cite{tobin}, could have the unintended consequence of propelling financial bubbles, such as those which have plagued international financial markets in the recent decades.

\appendix

\section{Information structures and information efficiency}
\label{infoapp}

Form the information theoretic point of view, the content of the signal $k_i^\omega$ can be quantified in one bit. Indeed, the entropy of the unconditional distribution over states, which is $\log\Omega$, is reduced to $\log (\Omega/2)$ by the knowledge of the signal $k_i^\omega$. Hence, the information gain is $\log 2$, i.e. one bit. This information gain allows agent $i$ to discriminate between two different conditional distributions of returns, whose means $E[R^\omega|k_i^\omega=\pm 1]$ are separated by an amount of order $1/N$. Indeed, for any two states $\omega$ and $\omega'$, by assumption $R^\omega-R^{\omega'}\sim 1/\sqrt{N}$. Now take the average over the states $\omega$ and $\omega'$ such that $k_i^\omega=+1$ and $k_i^{\omega'}=-1$ respectively. Given that the expected value of $R^\omega$ and $R^{\omega'}$ are the same, one finds that the average of the difference is of the order of the standard deviation of $R^\omega$, times the square root of the number $\Omega/2$ of samples, i.e. 
$$\left|E[R^\omega|k_i^\omega=+ 1]-E[R^\omega|k_i^\omega=- 1]\right|\sim s/\sqrt{N\Omega/2}\sim 1/N.$$
This difference is of the same order of the contribution of agents to the price $p^\omega$, hence it allows to differentiate meaningfully their investments $z_i^m$, depending on the signal they receive.

Let us now discuss market efficiency. A market is
efficient with respect to an information set if the public revelation
of that information would not change the prices of the securities \cite{fama}. This means that the best prediction of future returns (or prices), conditional on the information set, are present (discounted) prices.
{\it Strong efficiency} refers to the case where the information set includes the information
available to any of the participants in the market, including private
information. 

In our case, an agent who knew simultaneously the signals $k_i^\omega$ of all agents would be
able to know the state $\omega$, with probability one, for $\Omega\propto N$ and $N\to\infty$.

Indeed let $N_=$ be the number of pair of states $\omega$ and $\omega'$ which cannot be distinguished on the basis of the knowledge of all signals. For such pair of states, $k_i^\omega=k_i^{\omega'}$ must hold for all $i$, because otherwise there would be a signal $k_i^\omega\neq k_i^{\omega'}$ which allows to distinguish $\omega$ from $\omega'$. The probability $P\{N_=>0\}$ that there are at least two states
$\omega$ and $\omega'$ with different returns $R^\omega\neq R^{\omega'}$, but which cannot be distinguished given the signals, is upper bounded by the expected value of $N_=$. The latter can be easily evaluated, since for each pair of states the probability of them not being distinguishable is $P\{k_i^\omega=k_i^{\omega'},\forall i\}=2^{-N}$. The number of pairs is $\Omega(\Omega-1)/2$ so that 
$$P\{N_=>0\}\le E[N_=]=\Omega(\Omega-1)2^{-(N+1)},$$
and this vanishes for $N\to\infty$ in the case $\Omega\propto N$ we consider here.
We conclude that, if all signals were revealed, agents would be able to know which state $\omega$ has materialized. In this case, prices would not change only if  $p^\omega=R^\omega$ for all states $\omega$. 
Hence $H=0$ is equivalent to the strong form of information efficiency
\cite{efficiency}. 

\section{The statistical mechanics analysis}
\label{calc}
The competitive equilibrium solution of our problem can be  obtained through the minimisation of the following Hamiltonian function
\be
H_{\epsilon}=\frac{N^2}{4\Omega}\sum_{\omega,k_0} (R^{\omega} - p^{\omega, k_0})^2+\frac{\epsilon}{2}\sum_{i,m} z_i^m,
\ee
with $p^{\omega, k_0}=\frac{1}{N}\sum_{i,m}z_i^m\delta_{k_i^{\omega},m}+\sum_{k_0}\frac{z_0^{k_0}}{N}$.
In order to compute the minima of $H$ we introduce the partition function
\be
Z(\beta)=\int_0^{\infty} dz_0^+\int_0^{\infty} dz_0^-\cdots\int_0^{\infty} dz_N^+\int_0^{\infty} dz_N^- e^{-\beta H_{\epsilon}{\{z_i^m\}}} .
\ee
In the limit $\beta\to\infty$ integrals are dominated by those configurations $\{z_i^m\}$ that minimise the Hamiltonian.
The central quantity to compute is the free energy $f_{\beta}=-\beta^{-1}\log Z(\beta)$, which has to be averaged over the realisations of the disorder, namely $\{k_i^{\omega},R^{\omega}\}$. 
 In the following we are going to consider $k_i^{\omega}=\pm1$ with equal probability $\forall$ $i$, $\omega$, and we take $R^{\omega}=R+\frac{\tilde{R}}{\sqrt{N}}$, where $\tilde{R}$ are Gaussian variables with zero mean and variance equal to $s^{2}$.
In order to compute the average over the disorder $\avg{ f_{\beta}}$  we can resort to the so called replica trick through the identity $\log Z={\rm lim}_{M\to 0} (Z^M-1)/M$. The problem reduces then to that of computing the average over the disorder of the partition function of $M$ non interacting replicas of the system:
 \beas
 \avg{Z^{M}}&=&\Big\langle {\rm Tr}_{\{z\}}\prod_a\delta\left(N\overline{R}-\sum_i\overline{z}_{i,a}-\overline{z}_{0,a}\right) \times\\
 & & e^{-\beta\left[\sum_{a,\omega, k0}(NR^{\omega}-\sum_{i,m}z_i^m\delta_{k_i^{\omega},m}-z_0^{k0})^2+\epsilon\sum_{i,a}\frac{z_{i,a}^++z_{i,a}^-}{2} \right]}\Big\rangle,
 \eeas
 with $a\in\{1, \ldots, M\}$, $i\in\{1, \ldots, N\}$, $\omega\in\{1, \ldots, \Omega\}$, $m\in \{-1,1\}$ and $k0\in\{-1,1\}$ and $\overline{z}_{i.a}= (z_{i,a}^++z_{i,a}^-)/2$ . 
We verified  through numerical simulations that, for the specific public signal  $k_0$ that we considered in this paper, $\avg{z_0^+}=\avg{z_0^-}$ so, in order to simplify the calculation, we make the assumption $z_0^+=z_0^-= z_0$.
After performing a Hubbard-Stratonovich transformation in order to linearize the quadratic term of the Hamiltonian, taking the average over the quenched variables introduces an effective interaction between replicas:
 \beas
 \avg{Z^{n}} &=& \Big\langle \int \{d Q_{a,b}\}\{d \hat{Q}_{a,b}\}\{d \hat{R}\}{\rm Tr}_{\{z\}}\\
 & &e^{-\sum_{a,b}\hat{Q}_{a,b}\left(NQ_{a,b}-\sum_i \Delta_i^a\Delta_i^b\right) -\sum_a \hat{R}_a\left(N\overline{R}-\sum_i\overline{z}_{i,a}-z_{0,a}\right)} \times\\
 & & e^{-\beta N/\Omega\sum_{a,b,\omega}(\tilde{R}^{\omega})^2\left(\frac{\beta Q_{a,b}}{\alpha}+\delta_{a,b}\right)^{-1}
 -\beta\epsilon\sum_{i,a}\overline{z}_{i,a}}\times\\
 & & e^{ -\frac{\Omega}{2}{\rm Tr}\log\left(\frac{\beta Q_{a,b}}{\alpha}+\delta_{a,b}\right)}
 \Big\rangle,
\eeas
where we have introduced the overlap matrix $Q_{a,b}$ and the variables $\Delta_{i.a}= (z_{i,a}^+-z_{i,a}^-)/2$, while $\hat{Q}_{a,b}$ and $\hat{R}_a$ are conjugated variables that come from integral representations of $\delta$ functions:
\be
\delta(X-X_0)\propto \int  d \hat{X} e^{-\hat{X}(X-X_0)}.
\ee
In order to make further progress we consider the replica symmetric ansatz, namely we take
\bea 
Q_{a,b}&=&  q_0+\alpha\frac{\Phi}{\beta}\delta_{a,b}\\
\hat{Q}_{a,b} &=& -\frac{\beta^2\hat{q_0}}{\alpha^2}+\frac{\beta^2 \hat{q_0}/\alpha^2+\beta w/\alpha}{2}\delta_{a,b}
\eea
The resulting expression is handled in such a way to be able to use saddle point methods in the limit $N,\beta\to\infty$ (see \cite{CMV2009} for more details  on a similar calculation).
The final result is given in terms of the free energy
\be
f(q_0, \Phi, \hat{q}_0,w,\hat{R},z_0)= \frac{s^2+q_0}{1+\Phi}+ 2\frac{\hat{R}\overline{R}}{\alpha}-2\frac{\hat{R}z_0}{\alpha}+\frac{\Phi\hat{q_0}}{\alpha}-\frac{wq_0}{\alpha}+\frac{2}{\alpha}\Big\langle {\rm min}_{z\ge 0}\left\{V(z)\right\} \Big\rangle_t,
\ee
with the potential $V(z)$ given by
\be
V(z)=\frac{w}{2}\Delta^2 -\sqrt{\hat{q}_0} t\Delta-\hat{R}\overline{z}+\epsilon\overline{z}
\ee
and where we used $\langle\cdots \rangle_t$ to denote averages over the normal variable $t$.
The corresponding saddle point equations are
\bea
\label{w} w &=& \frac{\alpha}{1+\Phi}\\
\label{qh} \hat{q}_0&=& \frac{\alpha(s^2+q_0)}{(1+\Phi)^2}\\
\label{R} \overline{R} &=& z_0+\langle\Delta^*\rangle_t\\
\label{q} q_0 &=&  \langle{\Delta^*}^2\rangle_t\\
\label{chi}\Phi &=& \frac{\langle t\Delta^*\rangle_t}{\sqrt{\hat{q_0}}}\\
\label{Rh} \hat{R} &=& 0,
\eea
where
\bea
\Delta^*(t)&=&\theta(t-\tau)\frac{\sqrt{\hat{q}_0}}{w}(t-\tau)+\theta(-t-\tau)\frac{\sqrt{\hat{q}_0}}{w}(-t-\tau)\\
\tau&=&\frac{\epsilon}{\sqrt{\hat{q}_0}}.
\eea
Using these equations it is possible to compute $\langle H_{\epsilon}\rangle=\frac{q_0+s^2}{(1+\Phi)^2}$.
It is useful to define the three functions
\bea
\psi_r(\tau)&=&2\int_{\tau}^{\infty} dt e^{-t^2/2} (t-\tau)=\sqrt{\frac{2}{\pi}} e^{-\tau^2/2}-\tau {\rm erfc}\left(\frac{\tau}{\sqrt{2}}\right)\\
\psi_q(\tau)&=&2 \int_{\tau}^{\infty} dt e^{-t^2/2} (t-\tau)^2=(1+\tau^2) {\rm erfc}\left(\frac{\tau}{\sqrt{2}}\right)-\sqrt{\frac{2}{\pi}} \tau e^{-\tau^2/2}\\
\psi_{\Phi}(\tau)&=&2 \int_{\tau}^{\infty} dt e^{-t^2/2} t(t-\tau)={\rm erfc}\left(\frac{\tau}{\sqrt{2}}\right)\eea
It is now possible to express equations \eqref{R}, \eqref{q} and \eqref{chi} in terms of these non-linear functions.We can now look for a parametric solution in terms of $\tau$, and consider $\alpha$ as an independent variable.
From the definition of $\tau$ we have $\hat{q}_0=\epsilon^2/\tau^2$. Inserting equation \eqref{w} into equation \eqref{chi} we find 
\be \label{a}
\alpha=\frac{1+\Phi}{\Phi}\psi_{\Phi}(\tau),
\ee
while from equation \eqref{q} we get
\be
q_0=\frac{\epsilon^2}{\tau^2}\frac{\psi_{q}(\tau)}{\psi_{\Phi}^2(\tau)}\Phi^2.
\ee
Inserting these expressions into equation \eqref{qh} we obtain 
\be
\frac{\epsilon^2}{\tau^2}=\frac{s^2\psi_{\Phi}(\tau)}{\Phi(1+\Phi)}+\frac{\epsilon^2}{\tau^2}\frac{\psi_{q}(\tau)\Phi}{\psi_\Phi(\tau)(1+\Phi)},
\ee
from which
\be
\Phi_{\pm}=\frac{-1\pm\sqrt{1+4\psi_{\Phi}(\tau)s^2\frac{\tau^2}{\epsilon^2}\left(1-\frac{\psi_{q}(\tau)}{\psi_{\Phi}(\tau)}\right)}}{2(1-\psi_{q}(\tau)/\psi_{\Phi}(\tau))}.
\ee
Since $\Phi$ has the meaning of a distance between replicas the only physical solution is $\Phi=\Phi_+$.  Inserting this expression for $\Phi$ in the previous equations makes possible to express all order parameters and $\alpha$ in terms of the functions $\psi_r$, $\psi_q$, $\psi_{\Phi}$ and of the free parameters $\epsilon$ and $\tau$.\\
A parametric solution can be found also for the case of $\alpha$ fixed and $\epsilon$ variable.
From equation \eqref{chi} we find
\be
\Phi=\frac{\psi_{\Phi}(\tau)}{\alpha-\psi_{\Phi}(\tau)}.
\ee
From equation \eqref{q}
\be
q_0=\frac{\epsilon^2}{\tau^2}\frac{(1+\Phi)^2}{\alpha^2}\psi_{q}(\tau).
\ee
Finally, inserting this expression in equation \eqref{qh} we can now express $\epsilon$ as:
\be
\frac{\epsilon^2}{\tau^2}=\frac{\alpha s^2}{(1+\Phi)^2}\frac{1}{1-\frac{\psi_{q}(\tau)}{\alpha}}.
\ee
As before, using this expression, is now possible to  write the order parameters in terms of $\psi_r$, $\psi_q$, $\psi_{\Phi}$ and of the free parameters $\alpha$ and $\tau$.


\begin{thebibliography}{99}

\bibitem{Hart75} O. Hart (1975), 
{\em On the Optimality of Equilibrium When the Market Structure Is Incomplete}, Journal of Economic Theory  {\bf 11}, 418.

\bibitem{CassCitanna} D. Cass and A. Citanna (1998), 
 {\em Pareto Improving Financial Innovation in Incomplete Markets}, Economic Theory {\bf 11}, 467.
 
 \bibitem{MertonBodie} R. C. Merton and Z. Bodie (2005)
 {\em Design of Financial Systems: Towards a Synthesis of Function and Structure}, Journal of Investment Management {\bf 3}, 1. 
 
 \bibitem{Iceland} F. S. Mishkin, T. T. Herbertsson (2006), {\em Financial stability in Iceland}, Iceland Chamber of Commerce report, May 2006.
 
 \bibitem{Mishkin} F. S. Mishkin (1996). 
 {\em Understanding Financial Crises: A Developing Country Perspective}, Annual World Bank Conference on Development Economics,  29.

\bibitem{Jarrow} S. M. Turnbull, M. Crouhy and R. A. Jarrow (2008),  
Available at SSRN: http://ssrn.com/abstract=1112467
 
 
\bibitem{BHW2006}  W.A. Brock, C.H. Hommes, F.O.O. Wagener (2009),
 {\em More hedging instruments may destabilize markets} Journal of Economic Dynamics and Control {\bf 33}, 1912.

\bibitem{CMV2009} F. Caccioli, M. Marsili and P. Vivo (2009), 
{\em Eroding Market Stability by Proliferation of Financial Instruments}, European Physical Journal B {\bf 71}, 467.
 
\bibitem{MM2009} M.Marsili (2009), 
{\em Complexity and Financial Stability in a Large Random Economy} . Available at SSRN: http://ssrn.com/abstract=1415971. 
 
 \bibitem{fama} E. F. Fama (1970),  
{\em Efficient Capital Markets: A Review of Theory and Empirical Work},
Journal of Finance {\bf 25}, 383. 
 
\bibitem{GrossmanStiglitz} S. J. Grossman and J. E. Stiglitz (1980) 
 {\em On the Impossibility of Informationally Efficient
Markets} American Economic Review {\bf 70},  393.

\bibitem{Hommes} C. H. Hommes (2006), 
 Handbook of Computational Economics, in: L. Tesfatsion \& K.  L. Judd (eds.),  {\it Handbook of Computational Economics}  {\bf 2}, 1109 Elsevier.

\bibitem{LM99} T. Lux and M. Marchesi (1999), 
{\em Scaling and criticality in a stochastic multi-agent model of a financial market }, Nature  {\bf 397}, 498.

\bibitem{BMRZ} J. Berg, M. Marsili, A. Rustichini and R. Zecchina  (2001), 
{{\em Statistical mechanics of asset markets with private information}}, Quantitative Finance {\bf 1}, 203.

\bibitem{ShapleyShubik}  L. Shapley and M. Shubik (1997), 
{\em Trade Using One Commodity as a Means of
Payment},  Journal of Political Economy {\bf 85}, 937.

\bibitem{Pliska} S. R. Pliska (1997), {\em Introduction to Mathematical Finance: Discrete Time Models}, Blackwell, Oxford.

\bibitem{goldbaum} D. Goldbaum, (2006), {\em Self-organization and the persistence of noise in financial markets}, Journal of Economic Dynamics and Control, {\bf 30}, 1837.

\bibitem{tobin} M. Ul Haq, I. Kaul and I. Grunberg (eds) (1996), {\em The Tobin Tax: Coping With Financial Volatility}, Oxford University Press

\bibitem{efficiency} B. Malkiel (1992), {\em Efficient Market
Hypothesis}, in P. Newman, M. Milgate and J. Eatwell (eds.) {\it
New Palgrave Dictionary of Money and Finance}, MacMillan, London.  


\bibitem{hellwig} M. F. Hellwig, (1980), {\em On the aggregation of information in competitive markets}, Journal of Economic Theory, {\bf 22}, 477.

\end{thebibliography}
\end{document}